\documentclass[conference]{IEEEtran}
\IEEEoverridecommandlockouts
\usepackage{subcaption}
\usepackage[colorlinks=true,linkcolor=black,anchorcolor=black,citecolor=black,filecolor=black,menucolor=black,runcolor=black,urlcolor=black]{hyperref}
\usepackage{cite}
\usepackage{amsmath,amssymb,amsfonts}
\usepackage{graphicx}
\usepackage{textcomp}
\usepackage{xcolor}
\usepackage{amsmath}
\usepackage{algorithm}
\usepackage{algpseudocode}
\usepackage{pifont}
\usepackage{textcomp}
\usepackage{xcolor}
\usepackage[pscoord]{eso-pic}

\def\BibTeX{{\rm B\kern-.05em{\sc i\kern-.025em b}\kern-.08em
    T\kern-.1667em\lower.7ex\hbox{E}\kern-.125emX}}

\makeatletter
\let\old@ps@headings\ps@headings
\let\old@ps@IEEEtitlepagestyle\ps@IEEEtitlepagestyle

\def\confheader#1{%
  \def\ps@IEEEtitlepagestyle{%
    \old@ps@IEEEtitlepagestyle%
    \def\@oddhead{\strut\parbox[t]{\textwidth}{#1}\strut}%
    \def\@evenhead{\@oddhead}%
  }%
  \ps@IEEEtitlepagestyle%
}

\confheader{2024 6th International Conference on Electrical Engineering and Information \& Communication Technology (ICEEICT)  \\Military Institute of Science and Technology (MIST), Dhaka-1216, Bangladesh }

\makeatother

\newcommand{\placetextbox}[3]{
 \setbox0=\hbox{#3}
 \AddToShipoutPictureFG*{ \put(\LenToUnit{#1\paperwidth},\LenToUnit{#2\paperheight}){\vtop{{\null}\makebox[0pt][c]{#3}}}
 }
 }
 \placetextbox{.23}{0.055}{\small{979-8-3503-8577-9/24/\$31.00~\copyright 2024 IEEE}}

\makeatother

\makeatother

\title{ Transforming Precision: A Comparative Analysis of Vision Transformers, CNNs, and Traditional ML for Knee Osteoarthritis Severity Diagnosis\\
{}
}

\author{\IEEEauthorblockN{Tasnim Sakib Apon}
\IEEEauthorblockA{\textit{Dept. of CSE} \\
\textit{University of Maine System}\\
Orono, USA \\
sakibapon7@gmail.com}
\and
\IEEEauthorblockN{ Md.Fahim-Ul-Islam}
\IEEEauthorblockA{\textit{Dept. of CSE} \\
\textit{Brac University}\\
Dhaka, Bangladesh\\
fahim.islam@g.bracu.ac.bd}
\and
\IEEEauthorblockN{ Nafiz Imtiaz Rafin}
\IEEEauthorblockA{\textit{Dept. of CSE} \\
\textit{Brac University}\\
Dhaka, Bangladesh\\
nafiz.imtiaz@bracu.ac.bd}
\and
\IEEEauthorblockN{ Joya Akter}
\IEEEauthorblockA{\textit{Dept. of CSE} \\
\textit{Brac University}\\
Dhaka, Bangladesh \\
joya.akter@g.bracu.ac.bd}
\and
\IEEEauthorblockN{ Md. Golam Rabiul Alam}
\IEEEauthorblockA{\textit{Dept. of CSE} \\
\textit{Brac University}\\
Dhaka, Bangladesh \\
rabiul.alam@bracu.ac.bd}

}

\begin{document}






\maketitle

\begin{abstract}

Knee osteoarthritis is a degenerative joint disease that can cause severe pain and impairment. With increased prevalence, precise diagnosis by medical imaging analytics is crucial for appropriate illness management. This research investigates a comparative analysis between traditional machine learning techniques and new deep learning models for diagnosing knee osteoarthritis severity from X-ray pictures. This study does not introduce new architectural innovations but rather illuminates the robust applicability and comparative effectiveness of pre-existing ViT models in a medical imaging context, specifically for knee osteoarthritis severity diagnosis. The insights garnered from this comparative analysis advocate for the integration of advanced ViT models in clinical diagnostic workflows, potentially revolutionizing the precision and reliability of knee osteoarthritis assessments. This study does not introduce new architectural innovations but rather illuminates the robust applicability and comparative effectiveness of pre-existing ViT models in a medical imaging context, specifically for knee osteoarthritis severity diagnosis. The insights garnered from this comparative analysis advocate for the integration of advanced ViT models in clinical diagnostic workflows, potentially revolutionizing the precision and reliability of knee osteoarthritis assessments. The study utilizes an osteoarthritis dataset from the Osteoarthritis Initiative (OAI) comprising images with 5 severity categories and uneven class distribution. While classic machine learning models like GaussianNB and KNN struggle in feature extraction, Convolutional Neural Networks such as Inception-V3 and VGG-19 achieve better accuracy between 55-65\% by learning hierarchical visual patterns. However, Vision Transformer architectures like Da-VIT, GCViT and MaxViT emerge as indisputable champions, displaying  66.14\% accuracy, 0.703 precision, 0.614 recall, and AUC exceeding 0.835 thanks to self-attention processes. This analysis strongly promotes the deployment of sophisticated vision transformers over CNNs and traditional ML for increased precision in knee osteoarthritis diagnosis using X-ray picture categorization.

\end{abstract}


\section{Introduction}
Knee osteoarthritis (OA) is one of the most prevalent and disabling chronic joint diseases, affecting over 250 million people worldwide. It involves the degradation of articular knee cartilage and underlying bone, resulting in joint pain, stiffness, and impaired mobility. Osteoarthritis, a neurologically incurable disease that progresses and affects roughly 15
A significant technological barrier to the advancement of knee OA prevention and treatment is the lack of effective imaging characteristics to detect OA recurrence. \cite{b12}. The primary diagnostic method employed in clinical research appears to be medical imaging. By calculating the distance between the tibia and femur bones, the cartilage can be indirectly determined using standard X-ray imaging. Thus, cartilage cannot be accurately assessed with an X-ray. \cite{b20}. Due to this, a substantial amount of research has been done using various conventional machine-learning techniques to predict early osteoarthritis. Two examples of computer-aided techniques that have been introduced are dynamic contours and B-splines. \cite{b21}. Poor precision and dependability, as well as an inability to recognize minute cartilage changes, are some of these technologies' drawbacks. These methods may easily overlook the minute changes that take place in the early stages of knee osteoarthritis because osteoarthritis is a chronic condition that progresses slowly, on average by 2\% per year. Therefore, it is still difficult to create a reliable quantitative technique that is valid, reproducible, and change-sensitive. The outdated method may identify a cartilage thickness growth of only 2–3 percent per year. \cite{b13}. In osteoarthritis medical studies, body mass index, age, and gender are routinely used to identify people who are more likely to develop knee osteoarthritis. \cite{b15}. Since the effects and linkages of these markers are not fully understood, attempts to control these variables for early osteoarthritis identification and knee osteoarthritis progression have not been very successful. \cite{b16}, \cite{b17}.
This research provides a comparison analysis between classic machine learning models and current deep learning architectures for diagnosing knee OA severity from X-ray images. The study analyzes a dataset from the Osteoarthritis Initiative (OAI) comprising 1526 knee radiographs divided into 5 grades of increasing disease severity, with an uneven class distribution. This work strongly promotes the real-world adoption of vision transformers over CNNs and standard machine learning for exact knee OA severity detection utilizing automated image categorization. With future optimizations to model generalization, robustness, and inference time, ViT-based grading could successfully lead to appropriate non-surgical or surgical treatment selections to handle this expanding healthcare dilemma. The key contribution of the paper is given below:
\begin{itemize}
    \item We critically analyze the limitations of previously used techniques for knee OA categorization.
    \item By comparing classical ML models versus current deep learning methods for severity grading of our OA dataset.
    \item Furthermore, we advocate the adoption of vision transformers over CNNs and standard ML classifier models for better performance.
    \item We evaluate the performance by various performance metrics (accuracy, F1 score, precision, AUC score) where our ViT models outperform other models considering all these metrics.
    
\end{itemize}

The related works section critically analyzes the limitations of previously deployed techniques. The methodology section details the learning methods and proposed model. Accuracy, precision, recall and F1-score assessments follow using appropriate computations.

\section{Related Works}

\begin{table*}[!t]
 \caption{Overview of the Existing Related Studies}
    \begin{tabular}{@{}cccccc@{}}
        \hline
            Reference & Dataset & Preprocessing & Model & Outcome & Insight \\ 
        \hline
            PONGSAK et al {[}1{]} & OAI(1650 images) & Data Augmentation & \begin{tabular}[c]{@{}c@{}}YOLOv3\\ Detection\end{tabular} & Precision:0.850 F1:0.850              & \begin{tabular}[c]{@{}c@{}}Limited image usage,\\ but satisfactory \\ outcomes\\ were still gained\end{tabular}\\
        \hline
        Sudeep et al {[}2{]}      & OAI ( 4447 images)                                                             & \begin{tabular}[c]{@{}c@{}}Bounding Box \\ Annotation\end{tabular} & \begin{tabular}[c]{@{}c@{}}CNN\\ Regression\end{tabular}                   & \begin{tabular}[c]{@{}c@{}}Precision:0.73\\ Recall:0.73\\ F1-Score:0.73\\ Kappa:0.66\end{tabular}                                                                              & \begin{tabular}[c]{@{}c@{}}The model's\\  performance\\ improved significantly\\ after adjusting for \\ picture\\ quality differences\end{tabular} \\ \hline
        Joseph et al {[}3{]}      & \begin{tabular}[c]{@{}c@{}}OAI(8892 images),\\ MOST(3026 images)\end{tabular}  & Hist. equaliaztion                                                 & \begin{tabular}[c]{@{}c@{}}CNN-Reg\\ CNN-Reg*\end{tabular}                 & \begin{tabular}[c]{@{}c@{}}Classification loss: \\ Precision:0.43,\\ Recall:0.44,F1-\\ Score:0.43,\\ Regression loss:\\ Precision:0.61Recall:\\ 0.62F1-Score:0.59\end{tabular} & \begin{tabular}[c]{@{}c@{}}Template matching\\ had low accuracy,\\ CNNs\\ showed promise\end{tabular}                                              \\ \hline
        Abdelbasset et al {[}4{]} & OAI(1000 images)                                                               & Hist. equalization                                                 & \begin{tabular}[c]{@{}c@{}}Naive Bayes,\\ Random Forest\end{tabular}       & \begin{tabular}[c]{@{}c@{}}Accuracy:0.8298\\ Sensitivity:0.8715\\ Specificity:0.8065\end{tabular}                                                                              & \begin{tabular}[c]{@{}c@{}}Introduced approach\\ achieved superior\\ performance\end{tabular}                                                      \\ \hline
        Joseph et al {[}5{]}      & OAI(4446 images)                                                               & Localization                                                       & FCN,CNN                                                                    & \begin{tabular}[c]{@{}c@{}}Precision:0.61\\ Recall:0.63\\ F1-Score:0.61\end{tabular}                                                                                           & \begin{tabular}[c]{@{}c@{}}Suggested techniques\\ surpassed earlier methods\end{tabular}                                                           \\ \hline
        Alekei et el {[}6{]}      & \begin{tabular}[c]{@{}c@{}}OAI(4928 images),\\ MOST (3918 images)\end{tabular} & Standardlization                                                   & \begin{tabular}[c]{@{}c@{}}Multimodal\\ ML (LR,GBM,\\ LR,CNN)\end{tabular} & \begin{tabular}[c]{@{}c@{}}AUC:0.80,\\ AP:0.62\end{tabular}                                                                                                                    & \begin{tabular}[c]{@{}c@{}}Deep learning approach\\ shows promise for \\ prediction\end{tabular}                                                   \\ \hline
        Egor et el {[}7{]}        & OAI(4866 imges)                                                                & Augmentation                                                       & Various Models                                                             & \begin{tabular}[c]{@{}c@{}}Average Precision:0.55-0.61,\\ ROC AUC: 0.77-0.79\end{tabular}                                                                                      & \begin{tabular}[c]{@{}c@{}}Proposed transformer-\\ based method \\ for prognosis\end{tabular}                                                      \\ \hline
        Huy et al {[}8{]}         & OAI(4866 imges)                                                                & Various Preprocessing                                              & \begin{tabular}[c]{@{}c@{}}FCN,GRU,\\ LSTM,MMTF,\\ CLIMAT\end{tabular}        & \begin{tabular}[c]{@{}c@{}}FCN: RMSE(0.798-0.802)\\ GRU: RMSE(0.885-0.875)\\ LSTM:RMSE(0.888-0.874)\\ MMTF:RMSE(0.886-0.874)\\ CLIMAT:RMSE(0.812-0.808)\end{tabular}           & \begin{tabular}[c]{@{}c@{}}Proposed transformer-based\\ method \\ for prognosis\end{tabular}                                                       \\ \hline
    \end{tabular}
     \label{lit}
\end{table*}


    In persons over the age of 60, Osteoarthritis disease is a chronic cause of impairment. Osteoarthritis of the knee is a common condition marked by cartilage deterioration. A study aims at establishing a new automatic segmentation method for MRI-based examination of human knee cartilage thickness. The exam consisted of a double echo steady state (DESS) pattern, which compares cartilage and soft tissues, such as the synovial fluid, utilizing a 3T scanner and a knee coil. The approach was constructed using MRI 3-D scans in which an autonomous segmentation method divided the bone-cartilage interaction for the femur and tibia, yielding a descriptive area of the interface. The MR images are first resampled in the vicinity of the surface of the bone. Second, the cartilage is distinguished as a bright and uniform tissue utilizing texture-analysis techniques enhanced by filtering. The exterior limit of the cartilage can be detected using this procedure of omitting soft tissues. Third, a Bayesian decision criterion-based technology allows for the automatic separation of cartilage and synovial fluid. Finally, the developed technology was used to assess the cartilage thickness and variations in thickness for an individual across sessions. \par

    The slow degradation of articular cartilage is a symptom of osteoarthritis (OA). In a research method, the cartilage is segmented using a pixel-based segmentation process.   The study uses MATLAB R2013a to analyze 15 images, including regular and OA-affected ones. The ambient noise is minimized using a crude mask, and GLCM feature extraction is applied for segmentation.The accuracy of the classification of the individual into regular and OA-affected was determined to be 86.66 percent. \par

    Subsequently, learning-based Computer-Aided Diagnosis (CAD) added various features can be potential for enhancing knee OA diagnostic performance. Learning discriminative features, on the other hand, can be difficult,  when working with complicated data like X-ray pictures, which are commonly used to diagnose knee OA. One work presents a Discriminative Regularized Auto Encoder (DRAE) that enables to discover simultaneously relevant and discriminative features that cheval classiﬁcation accuracy. Very particularly, the usual Auto-Encoder learning objective is paired with a penalty concept called discriminative loss. This supplementary phrase seeks to drive discriminatory features into the learned model. The purpose of this study was to differentiate among definitive absence (KL grade 0) and premature OA presence (KL-G1 and KL-G2). The study evaluated five learning algorithms: SVM-RBF, LDA, SMC, and KNN, using 3900 knee joint pictures. The SVM-RBF classifier outperformed the others with a max-voting efficiency of 82.53 percent, F-measure efficacy of 83.48 percent, and 88.23 percent precision and 79.22 percent recall, demonstrating superior performance. 

   A study using T2Map MRI and Density-weighted Protons sequence identified osteoarthritis in knee articular structures. However, these diagnostics have poor responsiveness. A random forest approach was used to classify osteoarthritis into three severity classifications: A, B, and C. The model accurately predicted osteoarthritis data by 86.96 percent. Results from 33 patients at Indonesia's Cipto Mangunkusumo National Hospital showed the multiclass classification strategy was effective.Following the determination of the classification method, the model classifier is trained utilizing training data. The confusion matrix is used to evaluate and validate this classifier model. \par

  The study introduces a machine learning-based diagnostics method for knee osteoarthritis severity, classifying patients into distinct stages based on cartilage thickness. The method outperforms conventional methods with 97\% accuracy. \par

A new method for detecting radiographic osteoarthritis in knee X-ray pictures uses Kellgren–Lawrence classification ratings. The system uses subjectively graded X-rays to identify different phases of OA intensity. Over 95\% of mild OA patients were correctly diagnosed, but due to the significant difficulties that accompanied KL grades 4 (severe Osteoarthritis) and 5 (knee replacement), these stages were left from this research. Furthermore, this research was carried out as part of a longitudinal aging experiment that will allow imaging findings to be compared not only to medical Osteoarthritis aspects such as pain, but also to physiological measurements related to aging body functions that may contribute to Osteoarthritis severity.. \par

 Table \ref{lit} has been prepared to present an overview of the strategies that have been employed for accurate diseased  Knee Osteoarthritis categorization.

\section{Methodology}  \label{methodology}

\begin{figure*}[t]
    \centering
    \includegraphics[scale=0.34]{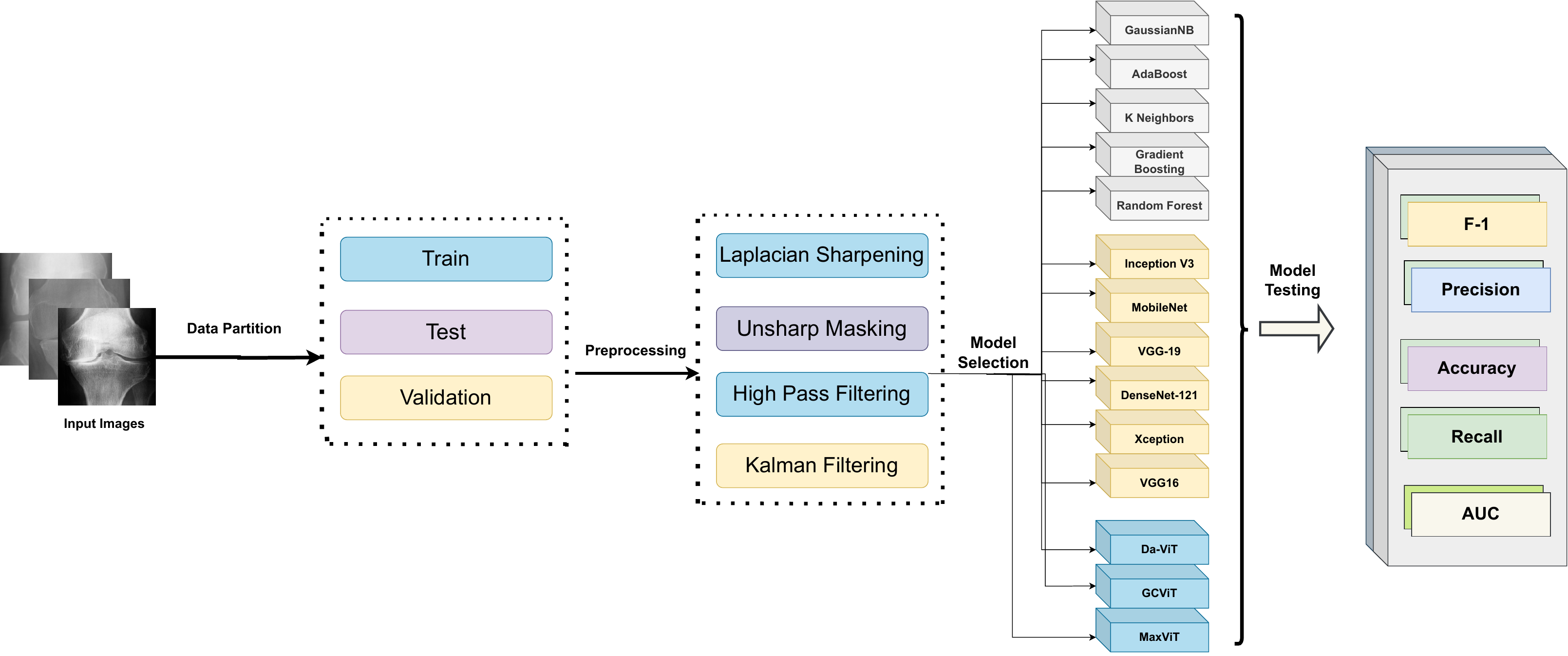}
    \caption{Flow diagram of classical machine learning models and vision transformers for diagnosing knee osteoarthritis severity from X-ray images advocating for the adoption of vision transformers, highlighting their potential to improve automated image categorization and guide treatment decisions with the help of image preprocessing methods}
    \label{flow-diagram}
\end{figure*}

Traditional approaches have limitations, necessitating a comparison of traditional machine learning with deep learning, notably vision transformers. The study recommends vision transformers (Da-VIT, GCViT, and MaxViT) for accurate OA severity assessment after analyzing 1526 knee radiographs as shown in Figure \ref{flow-diagram}. The paper examines past methodologies, describes the process, and assesses the suggested strategy using criteria such as accuracy and precision.

    \subsection{Data Acquisition and Preparation}
    
        \begin{table*}[!ht]
            \begin{center}
                \caption{Description and Distribution of Data by Classes.}
                \begin{tabular}{ c c c c c c  } 
                \hline
                 Knee Osteoarthritis X-ray Label & Description & Available & Training & Testing  & Validation \\
                 \hline
                 Grade 0:  & Healthy knee image. & 3253 & 2286 & 639  & 328\\
                 \hline
                 Grade 1 (Doubtful): & Doubtful joint narrowing with possible osteophytic lipping & 1495 & 1046 & 296  & 153 \\ 
                 \hline
                 Grade 2 (Minimal): & Definite presence of osteophytes and possible joint space narrowing & 2175 & 1516 & 447 & 212\\ 
                 \hline
                 Grade 3 (Moderate): & Multiple osteophytes, definite joint space narrowing, with mild sclerosis. & 1086 & 757 & 223  & 106 \\
                 \hline
                 Grade 4 (Severe): & Large osteophytes, significant joint narrowing, and severe sclerosis & 251 & 173 & 51 & 27\\
                  \hline
                \end{tabular}
            \end{center}
        \end{table*} 
        
        \subsubsection{Data Acquisition} According to radiographic data, more than half of American people over 65 had osteoarthritis disease around one or both joints in 2000. The osteoarthritis syndrome will be more common in 20 percent or more of US citizens by 2030, which would have a significant socioeconomic impact \cite{b16}, \cite{b17}. By the year 2050, Osteoarthritis and other reactive arthritis illnesses are predicted to affect at least 130 million people worldwide \cite{b14}. We have trained our model with “Knee Osteoarthritis Severity Grading Dataset” which was formulated in 2018. The Osteoarthritis Initiative (OAI), a multi-center, longitudinal, prospective observational research of knee osteoarthritis (OA) with the goal of identifying biomarkers for OA development and progression, provided the knee X-ray images used in our dataset. 
    
        \subsubsection{Data Description}
        Degeneration of the articular cartilage, a flexible, slick substance that typically shields bones from joint friction and impact, is what constitutes knee osteoarthritis. The disorder can also damage neighboring soft tissues and results in alterations to the bone that lies beneath the cartilage. By far the most prevalent form of arthritis to result in knee discomfort is knee osteoarthritis, which is frequently referred to as just knee arthritis. Rheumatoid arthritis, reactive arthritis, and many other uncommon kinds of arthritis can also hurt the knees.
        Numerous individuals who have computed tomography corroboration of Osteoarthritis do not demonstrate any symptoms, and the level of radiological alteration differs from person to person, making it challenging to identify Osteoarthritis in its initial stages. Measurement of hyaline cartilage alteration, that is also used to determine the feasibility of clinical therapies, is the primary technique for determining the structural development of Osteoarthritis. In this dataset, there are 4796 people whose ages range from 45 to 79. Additionally, there are 8260 knee joints in 4130 X-ray photos. This dataset contains 9786 images. The dataset was divided into 5 grading groups where grade 0 contains 604 x ray images, grade 1, grade 2, grade 3, grade 4 contains 275, 403, 200, 44 x ray images respectively. We have an uneven class distribution in our dataset that is partially imbalanced.

        \subsubsection{Image Preprocesing}
        We utilize a series of advanced approaches in this image preparation pipeline to improve the visual properties of disease images for better downstream analysis. Initially, a Laplacian sharpening operation is used to highlight minute features and edges in the photos, improving overall clarity. Following that, a high-pass filtering phase is applied using a specially constructed kernel to emphasize high-frequency components, hence improving image characteristics even further. Following that, a Kalman filter is used to smooth the images, utilizing a cutting-edge recursive method that efficiently minimizes noise and temporal changes in the images. To enhance its effectiveness in the context of image smoothing, the Kalman filter is designed with carefully chosen parameters, including measurement and process noise covariances. This extensive preprocessing pipeline is applied systematically to a broad variety of pictures sorted into separate segments and classes, resulting in visually polished outputs. These methods target both fine detail enhancement and noise reduction, contributing to overall image quality improvement for our classification assignment.
     
    \subsection{Model Specification}

    A wide range of machine learning techniques for image categorization are included in the model specifications. To make use of classical techniques, five classic machine learning models—GaussianNB, AdaBoost, K Neighbors, Gradient Boosting, and Random Forest—are used. Six Convolutional Neural Networks (CNNs) that illustrate the application of deep learning techniques are also included in the model repertoire. These include Inception-V3, MobileNet, VGG-19, DenseNet-121, Xception, and VGG-16. Moreover, the model specifications present the incorporation of the most advanced image classification transformers, namely Da-VIT, GCViT, and MaxViT, which are prime examples of the utilization of state-of-the-art vision transformers to improve efficiency and precision for classifying diseased knee images. The GCViT improves picture classification by using global context self-attention efficiently, reducing inductive bias, and obtaining cutting-edge results without pre-training. The DaViT here resides in its innovative use of dual self-attention mechanisms—spatial tokens and channel tokens—effectively capturing both global context and local interactions, resulting in state-of-the-art performance with efficient calculations.Besiedes, MaxViT overcomes the scalability limitation of self-attention in vision transformers by introducing an efficient and scalable multi-axis attention model that allows global-local spatial interactions on arbitrary input resolutions with linear complexity, resulting in improved image classification task performance \cite{gcvit,davit,maxvit}.

\subsection{Models of Vision Transformers}

\textbf{Configurations of Da-VIT, GCViT, and MaxViT:}
\begin{itemize} \item \textbf{Da-VIT:} The configuration has 12 layers, a hidden size of 768, 12 attention heads, and a patch size of 16. The decision was driven by the dual attention mechanism's ability to effectively capture subtle characteristics in knee X-ray images across both spatial and channel dimensions.
    \item \textbf{GCViT:} The model has 24 layers, a dimension of 1024, 16 attention heads, and a global context window size of 14. The global context-enhancing self-attention mechanism developed by GCViT is well-suited for medical datasets such as OAI, which consist of diverse visual characteristics and artifacts.
    \item \textbf{MaxViT:} The model incorporates a multi-axis attention mechanism, with 32 transformer blocks and a model width of 512. MaxViT was selected for its efficiency in processing high-resolution pictures and maintaining both local and global image features, making it perfect for extensive knee image studies.
\end{itemize}

\textbf{Hyperparameters:} \begin{itemize} \item \textbf{Learning Rate:} A learning rate of 1e-4 was established for all models, based on grid search optimization findings for optimal convergence.
    \item \textbf{Batch Size:} Chosen as 16 to balance computational efficiency and model performance within GPU memory limits.
    \item \textbf{Optimizer:} AdamW optimizer was utilized with a weight decay of 0.01 for its effectiveness in sparse gradient management.
\end{itemize}

\textbf{Data Augmentation:}
Data augmentation techniques such as rotation, zoom, and horizontal flipping were utilized to introduce robustness to differences in knee X-ray orientations and scales.

\subsection{Comparative Analysis using CNNs and Traditional ML}

\textbf{CNN Architectures:}
Inception-V3 and VGG-19 were configured with standard parameters, chosen for their demonstrated performance in image classification, giving a benchmark for ViT's evaluation.

\textbf{Traditional ML Models:}
Models like Random Forest and Gradient Boosting were utilized with default scikit-learn parameters to provide a complete comparison across machine learning approaches, showcasing the gains enabled by deep learning in medical imaging.

\subsection{Rationale Behind Model Selection}

The selection of deep learning architectures was informed by their demonstrated effectiveness in various image identification challenges. Each model's particular approach to image processing is vital for negotiating the complexity inherent in medical images, which are characterized by small variances and significant variability. The combination of classic ML models and proven CNNs allows for a contextualized evaluation of ViT models' achievements in knee osteoarthritis severity diagnosis using X-ray images.

\section{Result Analysis}

\begin{table}[!t]
    \begin{center}
        \caption{ Performance Evaluations Comparison of Models on Knee Osteoarthritis Severity Diagnosis } 
        \begin{tabular}{c c c c c c} 
            \hline
            \multicolumn{6}{c}{Performance Assesment} \\
            \hline
            Model & Acc(\%) & Precision & Recall & F-1 & AUC \\ 
            \hline
            \hline
            GaussianNB & 21.00 & 0.257 & 0.262 & 0.187 & 0.555 \\
            AdaBoost & 37.50 & 0.238 & 0.234 & 0.201 & 0.604 \\
            K Neighbors & 38.37 & 0.487 & 0.263 & 0.271 & 0.572 \\
            Gradient Boosting & 39.95 & 0.291 & 0.231 & 0.177 & 0.629 \\
            Random Forest & 41.52 & 0.201 & 0.237 & 0.196 & 0.686 \\ 
            \hline
            Inception-V3 & 60.41 & 0.663 & 0.483 & 0.434 & 0.872 \\
            MobileNet & 56.05 & 0.625& 0.418 & 0.378 & 0.859 \\ 
            VGG-19 & 65.25 & 0.708 & 0.577 & 0.504 & 0.903 \\
            DenseNet-121 & 55.81 & 0.622 & 0.464 & 0.373 & 0.858 \\
            Xception & 57.38 & 0.644 & 0.463 & 0.383 & 0.846 \\
            VGG-16 & 60.65 & 0.650 & 0.510 & 0.457 & 0.882 \\
            \hline
            Da-VIT & 65.35 & 0.706 & 0.606 & 0.603 & 0.848 \\   
            GCViT & 66.14 & 0.703 & 0.598 & 0.618 & 0.835 \\ 
            MaxViT & 66.14 & 0.666 & 0.614 & 0.618 & 0.860 \\
            \hline
        \end{tabular}
        \label{ML_Model_scores}
    \end{center}
\end{table}

The full performance evaluation in Table \ref{ML_Model_scores} provides a deep understanding of several picture classification models. Traditional machine learning models such as GaussianNB, AdaBoost, K Neighbors, Gradient Boosting, and Random Forest perform poorly in terms of accuracy, with GaussianNB trailing at 21.00\%, highlighting challenges in capturing detailed visual data. Convolutional Neural Networks (CNNs) such as Inception-V3, MobileNet, VGG-19, DenseNet-121, Xception, and VGG-16, on the other hand, show significant increases, with accuracy rates ranging from 55.81\% to 65.25\% when exploiting deep hierarchical features.

The standout performances, however, are the cutting-edge Vision Transformer (ViT) models Da-VIT, GCViT, and MaxViT. Da-VIT outperforms CNNs like VGG-19 and DenseNet-121 with an astounding accuracy of 65.35\%. GCViT and MaxViT significantly improve performance with accuracies of 66.14\%, demonstrating the constancy of ViT models in outperforming classical and CNN models.ViT models consistently outperform other models in terms of precision, recall, and F-1. Da-VIT, for example, has a precision of 0.706 and a recall of 0.606, demonstrating its ability to correctly classify positive cases. Precision-recall balances are demonstrated by GCViT and MaxViT, proving their powerful categorization capabilities.

The superiority of the ViT models is further highlighted by the AUC statistic, where Da-VIT, GCViT, and MaxViT routinely score above 0.835. This demonstrates their ability to discern across different classes and cements their status as leaders in picture classification challenges. Finally, the numerical analysis demonstrates that Vision Transformer models outperform classic machine learning models and CNNs in terms of not only accuracy but also precision, recall, and AUC values.

\section{Conclusion} \label{conclusion}
This investigation of detecting knee osteoarthritis severity from X-ray pictures shows compelling evidence for vision transformer models over traditional machine learning and convolutional neural networks. Utilizing a multi-center dataset of 1526 knee radiographs categorized into 5 progressive disease categories, advanced ViT architectures including DaVIT, GCViT and MaxViT achieve exceptional 66\% grading accuracy, 0.7 precision, 0.6 recall and AUC beyond 0.83. In comparison, conventional models like as GaussianNB obtain only 21\% accuracy, hampered by problems in informative visual feature extraction from scans. While CNNs like Inception-V3 achieve advances exceeding 65\% accuracy via hierarchical feature learning, their performance lags behind the latest vision transformers tuned for both local and global context modeling. We strongly encourage priority application of ViT models in real-world automated systems for robust knee osteoarthritis evaluation to guide suitable clinical decisions. Their advanced self-attention systems can record small visual patterns to advance diagnostics and monitoring. Further enhancements to transformer model generality, efficiency and clinical integration can provide a pathway for AI-assisted severity rating that improves conservative and surgical care of this crippling, chronic illness affecting over 250 million patients worldwide. Intelligent knee OA image categorization with transformers has the potential to radically improve orthopedic therapy. Table \ref{ML_Model_scores} depicts the evaluation scores of different models.

\end{document}